\documentclass[twocolumn,prd,nofootinbib,preprintnumbers,tightenlines,superscriptaddress]{revtex4}
\pdfoutput=1
\usepackage{amsmath,amssymb,amsfonts}
\usepackage{graphicx}
\usepackage[normalem]{ulem}
\usepackage{slashed}
\usepackage{color}
\usepackage{physics}
\usepackage[
	colorlinks=true,
	citecolor=black,
	linkcolor=black,
	urlcolor=blue,
	hypertexnames=false]{hyperref}
\usepackage{cleveref}

\newcommand{\colvec}[1]{\left(\begin{array}{c}#1\end{array}\right)}

\newcommand{\colmatt}[1]{\left(\begin{array}{cc}#1\end{array}\right)}

\newcommand{\beq}{\begin{equation}}
\newcommand{\eeq}{\end{equation}}

\definecolor{darkgreen}{rgb}{0.0, 0.5, 0.0}

\begin{document}

\title{The Seiberg-Witten Axion}

\author{Csaba Cs\'aki}
\email{csaki@cornell.edu}
\affiliation{Department of Physics, LEPP, Cornell University, Ithaca, NY 14853, USA}

\author{Rotem Ovadia}
\email{rotemov@gmail.com}
\affiliation{Department of Physics, LEPP, Cornell University, Ithaca, NY 14853, USA}
\affiliation{Racah Institute of Physics, Hebrew University of Jerusalem, Jerusalem 91904, Israel}

\author{Maximilian Ruhdorfer}
\email{m.ruhdorfer@stanford.edu}
\affiliation{Department of Physics, LEPP, Cornell University, Ithaca, NY 14853, USA}    
\affiliation{Stanford Institute for Theoretical Physics, Stanford University, Stanford, CA 94305, USA}

\author{Ofri Telem}
\email{t10ofrit@gmail.com}
\affiliation{Racah Institute of Physics, Hebrew University of Jerusalem, Jerusalem 91904, Israel}

\author{John Terning}
\email{jterning@gmail.com}
\affiliation{QMAP, Department of Physics, University of California, Davis, CA 95616, USA}

\begin{abstract} 
We present a fully calculable UV complete toy model of a Peccei-Quinn (PQ) axion coupled to magnetic monopoles as well as electric charges. The theory has manifest electric-magnetic duality built in. We find that the  axion-photon coupling contains the usual anomaly term, plus periodic corrections which  can also become large if the monopole is light, without violating the discrete axion shift symmetry. These additional periodic terms can be identified as the non-perturbative corrections due to the monopoles (and other BPS states), but can also be interpreted as a sum over instanton corrections. The key aspect helping reconcile axion coupling quantization with electric-magnetic duality is the fact that the axion itself undergoes a non-linear transformation under electric-magnetic duality. The theory analyzed here is just the original ${\cal N}=2$ supersymmetric $SU(2)$ Seiberg-Witten theory, which contains a PQ axion due to an anomalous spontaneously broken global $\mathcal{R}$-symmetry, as well as massless fermionic monopoles and dyons at special points in the moduli space. Hence the entire machinery of the Seiberg-Witten solution can be applied to reliably calculate the photon-axion coupling in different duality frames. We show explicitly that the physically observable axion-photon amplitude is duality invariant, as it had to be. 
\end{abstract}

\maketitle

\section{Introduction}

QCD Axions are widely studied theoretically and experimentally, as they provide an elegant solution to the strong CP problem and may also be the main component of dark matter.  As a Goldstone boson of the anomalous Peccei-Quinn (PQ) symmetry, the axion is an angle and the action is invariant\footnote{Up to $2\pi n$ shifts.} under $a\rightarrow a+2\pi f$, where $f$ is the VEV of the PQ scalar. Correspondingly, the PQ anomaly-induced axion-photon coupling is widely thought to be quantized, which forms the basis of most experimental axion searches. 
A more general formulation, in which the axions are also coupled to magnetically charged particles, was explored in refs.~\cite{Sokolov:2021eaz,Sokolov:2022fvs,Sokolov:2023pos}. They found that the inclusion of fermionic monopoles charged under the PQ symmetry modifies the form of the axion-photon coupling and, naively, seems to destroy its quantization.  However, Ref.~\cite{Heidenreich:2023pbi} claimed that such a lack of quantization is internally inconsistent and proposed a fix for this apparent inconsistency, by ascribing a transformation rule for the field strength $F$ under $a\rightarrow a+2\pi f$. Unfortunately, the latter fix leads to the theory being strongly ruled out phenomenologically. 

In this paper, we present a fully calculable model of a PQ axion with fermionic magnetic monopoles carrying a non-vanishing PQ charge. This allows us to examine the form of the photon-axion coupling and to better understand the  dynamics generating these couplings. We find that the form of the axion coupling to $F_{\mu\nu}\widetilde{F}^{\mu\nu}$ contains additional dynamically generated periodic terms, while the IR EFT is manifestly covariant under electromagnetic (EM) duality. One key aspect that helps reconcile the seeming contradiction between duality and coupling quantization is the fact that the axion itself transforms non-linearly under the electric-magnetic duality. In the electric frame the contribution of the electric charges appear as the usual anomaly induced linear term, while the contribution of all magnetically charged states appear as non-perturbative effects (which can be identified with instantons). These non-perturbative effects give rise to periodic functions hence automatically invariant under the discrete axion shift symmetry. In the magnetic frame it is the contribution of the magnetic charges that gives the linear coupling of the {\it dual axion}, while the contribution of the electric charges appear as non-perturbative effects. As an ultimate test of duality we show that the canonically normalized axion-photon matrix elements are invariant (up to a phase) under duality transformations, as they have to be, since duality gives two different descriptions of the exact same physics.

Our axion model is none other than the original $\mathcal{N}=2$ $SU(2)$ gauge theory without fundamental matter studied by Seiberg and Witten (SW) in their seminal work \cite{Seiberg:1994rs}. Surprisingly, this theory is also a prime example for studying the properties of a strongly coupled axion in a calculable, UV complete theory. The theory has a global $U(1)_{\cal R}$ symmetry which is anomalous under the strongly coupled $SU(2)$ group. This theory has a moduli space of vacua, with the $SU(2)$ gauge group broken to $U(1)$ (``the Coulomb branch")---it is this $U(1)$ that will play the role of our standard $U(1)$ for EM. 
On this moduli space the $U(1)_{\cal R}$ global symmetry is also spontaneously broken, providing a Goldstone boson, the axion. Since this symmetry is also anomalous (both under the original $SU(2)$ gauge group and the remaining $U(1)$), it will have all the necessary properties to serve as a toy model for a PQ axion. Most importantly, it is well-known from the Seiberg-Witten solution that at a specific point in the moduli space a magnetic monopole hypermultiplet becomes massless (while at another point a dyon becomes massless). The monopole (and the dyon) also carry a non-vanishing $U(1)_{\cal R}$ charge, hence their masses can be thought of as fully arising from the spontaneous PQ breaking, providing a non-trivial example of an axion model with a light magnetic monopole, which will provide additional contributions to the photon-axion coupling.

The IR dynamics of this SW theory is described by $\mathcal{N}=2$ Super Quantum Electrodynanamics (SQED) which is manifestly covariant under EM $SL(2,Z)$ duality. In the electric frame the axion, $a$, is identified with the phase of the chiral superfield $A$ which completes the vector superfield $V$ to a full ${\cal N}=2$ vector multiplet, while in the magnetic frame the dual axion $a_D$ is the phase of $A_D$ which completes $V_D$. Since $A$ and $A_D$ are related to each other via a complicated non-linear transformation (determined by the SW solution) so is the axion $a$ and the dual axion $a_D$.

We will see that in the electric frame the axion $a$ couples as\\
\begin{eqnarray}\label{eq:axion coupling}
    &&-\frac{e^2}{16\pi^2 f}\,F_{\mu\nu}\widetilde{F}^{\mu\nu}\times\nonumber\\[5pt]
    &&\left\{N_a a -\sum_{k=1}^\infty  \left[b_k \sin \left(\frac{4ka}{f}\right)+ c_k \cos \left(\frac{4ka}{f}\right)\right]\right\}~,~~~
\end{eqnarray}
where $N_a$ is an integer and $b_k$ and $c_k$ are calculable from Seiberg-Witten (SW) theory. The first term is the familiar linear term from the perturbative PQ anomaly, while the second term is a non-perturbative contribution, which is fully calculable in SW theory. Similar periodic couplings arise for the ordinary QCD axion as well due its mixing with the pion, see for example~\cite{Agrawal:2017cmd,Agrawal:2023sbp}.\footnote{We thank Matt Reece for emphasizing this.} Its interpretation, as pointed out in the original paper and elaborated below, is as the sum of all quantum contributions from heavy BPS monopoles and dyons, which can also be interpreted as a sum of instanton corrections.\footnote{These non-perturbative corrections to the photon-axion coupling are the analogs of the non-perturbative corrections to the axion potential and axion mass first calculated in~\cite{Fan:2021ntg}. These corrections can also be underdstood in the language of generalized symmetries, as emerging from the UV breaking of a magnetic 1-form symmetry participating in a higher group with PQ symmetry \cite{Cordova:2022ieu,Cordova:2023her}.} The contributions from an individual BPS state to the anomaly were first calculated in ~\cite{Csaki:2010rv} using EM duality, and the explicit SW solution shows that they must sum up to the periodic terms in \eqref{eq:axion coupling}. 
The latter axion coupling is manifestly periodic under $a\rightarrow a+2\pi f$, up to the usual unphysical $2\pi N$ shift of the action. In the magnetic dual theory one obtains a similar expression for the coupling of the dual axion to the dual photons, where now the contribution of the magnetic monopole will be interpreted in terms of the perturbative anomaly, while the effects of the electrically charged objects will give another periodic contribution due to non-perturbative effects. Once the proper normalizations and the relation between the axion decay constant vs. the dual decay constant are identified one can show that physical matrix elements agree in the electric and magnetic descriptions. 

 We thus have a concrete proof-of-principal for axion electrodynamics which has
\begin{itemize}
    \item Additional periodic coupling terms \eqref{eq:axion coupling} generated non-perturbatively from BPS monopoles and dyons.
    \item Manifest EM duality, where the axions in different duality frames are non-linearly related to each other.
\end{itemize}
The properties of the SW axion studied in this paper vary continuously over the moduli space of the theory. At weak coupling the scale $f$ of spontaneous $R$-symmetry breaking is much larger than the scale $\Lambda$, in which the theory becomes strongly coupled. In this regime the SW axion resembles the QCD-axion, for which $\Lambda_{QCD}/f_{PQ}\ll 1$. As one moves closer to the strongly coupled regime, the scale separation between $f$ and $\Lambda$ is lost, and the SW axion becomes an analog of $\eta'$ state for QCD (with the caveat that supersymmetry still ensures that the axion remains massless). In particular, the corrections to the axion coupling become large around the monopole point where $f\lesssim\Lambda$. In this regime, we expect that an addition of supersymmetry breaking to result in an axion mass of the same order as the SUSY breaking scale. 
While our study focuses on a toy model and not a full-fledged BSM model, it does have two major direct implications for the search for real-world axions and axion-like particles (ALPs). The first is that the axion coupling need not be quantized in integer multiples of $e^2/16\pi^2$, and can even be \textit{large}, depending on the value of $b_k$. The second is that axion electrodynamics without any additional modifications can be manifestly covariant under EM duality, and does not suffer from the phenomenological deal-breakers outlined in \cite{Heidenreich:2023pbi}.  This also implies that the 
implementation of duality in the seemingly 
non-duality covariant equations ubiquitous in the literature \cite{Sikivie:1983ip, Raffelt:1987im} 
will have to be reexamined.

\section{Overview of Seiberg-Witten} 
%
The ${\mathcal N}=2$ supersymmetric $SU(2)$ gauge theory without fundamental matter fields, analyzed by Seiberg and Witten \cite{Seiberg:1994rs} 
provides a nice laboratory to explore axion couplings to mutually non-local charges\footnote{See \cite{Bilal:1996sk, Tachikawa:2013kta} for excellent reviews of Seiberg-Witten theory. Here we follow the conventions of  \cite{Tachikawa:2013kta}.}. 
The IR dynamics of Seiberg-Witten theory involves a continuum of degenerate $\mathcal{N}=2$ vacua parametrized by a complex coordinate $u$. In all of these vacua the $SU(2)$ gauge symmetry is Higgsed to $U(1)$ by the adjoint scalar $\phi$ in the $SU(2)$ gauge multiplet. For this reason, the continuum of vacua is also called the \textit{Coulomb branch}. The Higgsing gives masses to $W^\pm$ gauge bosons and their four chargino superpartners.

The moduli space of the Coulomb branch is parametrized by a complex coordinate $u \equiv \frac{1}{2}{\rm tr}(\phi^2)$, where $u \neq 0$.
The IR dynamics is given in terms of the photon\footnote{The $\mathcal{N}=2$ gauge multiplet of the photon consists of an $\mathcal{N}=1$ chiral superfield $A$ and an $\mathcal{N}=1$ vector superfield $V$, whose supersymmetric field strength is $W_\alpha$.} $(A,V)$ or, equivalently, the ``dual photon" $(A_D,V_D)$. The two are not separate degrees of freedom but rather nonlinear functions of each other, with the chiral superfield components related by
\begin{equation}~\label{eq:AAD}
    A_D=\frac{\partial {\cal F}(A)}{\partial A}~~,~~A=-\frac{\partial {\cal F}_D(A_D)}{\partial A_D}\,.
\end{equation}
Here ${\cal F}(A)$ is the exactly calculable \cite{Nekrasov:2002qd} \textit{Prepotential} and ${\cal F}_D$ is its exactly calculable Legendre transform, the \textit{dual Prepotential}. The VEVs $A^v(u)$ and $A^v_D(u)$ are also known explicitly for Seiberg-Witten theory, and are given by
\begin{eqnarray}~\label{eq:AADu}
    A^v(u)&=&\sqrt{u+2\Lambda^2}\,{}_2F_1\left(-\frac{1}{2},\frac{1}{2},1;\frac{4\Lambda^2}{u+2\Lambda^2}\right)\nonumber\\[5pt]
    A^v_D(u)&=&i\frac{u-2\Lambda^2}{2\Lambda}\,{}_2F_1\left(\frac{1}{2},\frac{1}{2},2;\frac{2\Lambda^2-u}{4\Lambda^2}\right)\,,
\end{eqnarray}
where ${}_2F_1$ is the \textit{Gauss hypergeometric function}. Note that $A(u)$ has branch points at $u=\pm 2\Lambda^2$, while $A_D(u)$ has branch point at $u=- 2\Lambda^2$. At these branch points, BPS states become exactly massless. From the monodromies of $(A_D(u),\,A(u))$ from \eqref{eq:AADu} around these points, we can read-off the charges of the BPS states that become massless at these points; at $u=2\Lambda^2$ this is a monopole hypermultiplet with EM charges $(g,q)=(1,0)$ (in the language of the weak coupling), while at $u=-2\Lambda^2$ they are dyons of charge $(g,q)=(1,2)$. Finally, the holomorphic gauge coupling on the moduli space $\tau$ is exactly computable on the Coulomb branch, and is given by
\begin{equation}\label{eq:tau}
    \tau=\pdv[2]{{\cal F}}{A} \equiv \frac{\theta}{2\pi} + \frac{4 \pi i}{e^2} \,.
\end{equation}
Note that $\tau$ is a nonlinear function of $A$ \textit{as an $\mathcal{N}=1$ chiral superfield}. To get the numerical values of $e$ and $\theta$ we have to substitute the VEV $A(u)$ from \eqref{eq:AADu}. The  low-energy physics is completely determined by ${\cal F}(A)$, namely 
\beq
    {\mathcal L}_{IR}=
    \frac{1}{8 \pi i} \int d^4 \theta \frac{\partial {\cal F}}{\partial A} {\overline A}+
     \frac{1}{8 \pi i}  \int d^2 \theta\, \tau(A)\,
     W^{\alpha}W_\alpha +h.c. \,.
     \label{eq:SWLagrangian}
\eeq
The action is canonically normalized by rescaling the fields as $A \rightarrow e A$. However, this should be done with caution since the prepotential is only holomorphic in the original fields.

We now briefly describe two important regions of the Coulomb branch, namely, weak coupling and the strong coupling singularities.
\\ \quad\\
\textit{Weak Coupling} At large $|u|$, the theory is Higgsed at weak coupling, and the Coulomb branch coordinate $u$ coincides with $\frac{1}{2}{\rm tr}\phi^2=A^2$. In this weak coupling regime, the spectrum of the theory consists of the massless photon multiplet $A$, two $W^{\pm}$ multiplets getting their mass from the adjoint VEV $u=\frac{1}{2}{\rm tr}\phi^2=A^2$, and a tower of heavy semiclassical 't Hooft-Polyakov monopoles and their dyonic excitations, all BPS states whose mass satisfies
\begin{eqnarray}\label{eq: BPS mass}
    m^{BPS}_{(g,q)}=|A^v(u)\,q+A^v_D(u)\, g|\,,
\end{eqnarray}
where $(g,q)$ are the magnetic and electric charges of every state in the tower.
\\ \quad\\
\textit{Strong Coupling} At $u \sim \Lambda^2$ the IR EFT, expressed in terms of the photon $A$, is strongly coupled. In the vicinity of the $u=2\Lambda^2$ singularity we can work with the dual complexified gauge coupling $\tau_D \equiv \frac{\theta_D}{2\pi} + \frac{4 \pi i}{e^2_{D}} $, given by $\tau_D=\partial^2 {\cal F}_D(A_D)/\partial A^2_D$ and related to $\tau$ via \textit{S-duality} $\tau_D=-1/\tau$. One can verify that $e_D$ runs logarithmically to $0$ at $u=2\Lambda^2$. In the neighborhood of $u=2\Lambda^2$ the charge $(1,0)$ \textit{monopole hypermultiplet} $(M,\widetilde{M})$ becomes light, and there is an additional Yukawa term in the superpotential,
\begin{eqnarray}\label{eq:monterm}
    W = A_D M \widetilde{M}\,.
\end{eqnarray}
Exactly at $u=2\Lambda^2$, $A_D=0$ and the monopoles become massless. In this regime, we can use an equivalent, \textit{S-dual} description of the same IR physics in terms of the dual photon $A_D$ and the dual prepotential ${\cal F}_D(A_D)$, namely
\begin{eqnarray}
    {\mathcal L}^D_{IR}=&&\frac{1}{8 \pi i} \int d^4 \theta \frac{\partial {\cal F}_D}{\partial A_D} {\overline A_D}+\nonumber\\[5pt]
    &&
     \frac{1}{8 \pi i}  \int d^2 \theta\, \tau_D(A_D)\,
     W^{\alpha}_DW_{D\alpha} +h.c.\,,
     \label{eq:SWLagrangianD}
\end{eqnarray}
in addition to the monopole term \eqref{eq:monterm}.

The \emph{exact} expressions for the prepotential ${\cal F}(A)$ and the dual prepotential ${\cal F}_D(A_D)$ (which is the Legendre transform of ${\cal F}(A)$) are given by \cite{Tachikawa:2013kta, Nekrasov:2002qd, Seiberg:1994rs, Ito:1995ga, Chan:1999gj}
\begin{eqnarray}\label{eq:prepotential}
    &&{\cal F}(A)= \frac{1}{2\pi i} \left\{-4A^2 \left(\ln \frac{2A}{\Lambda}-\frac{3}{2}\right)+ A^2 \sum_{k=1}^\infty d_k \left(\frac{\Lambda}{A}\right)^{4k}\right\}\nonumber\\[5pt]
    &&{\cal F}_D(A_D)= \frac{1}{4\pi i}\left\{ A_D^2 \left(\ln \frac{-iA_D}{32\Lambda}-\frac{3}{2}\right)-16i A_D+\right.\nonumber\\[5pt]
    &&~~~~~~~~~~~~~~~~~~~~~~~~~~~~~~~~~~\left.A_D^2 \sum_{k=1}^\infty d^D_k \left(\frac{A_D}{\Lambda}\right)^{k} \right\}~.
\end{eqnarray}
The first term in ${\cal F}(A)$ corresponds to a one-loop contribution from the massive gauge multiplets, while the first term in ${\cal F}_D(A_D)$ corresponds to the one-loop contribution from the light monopoles $M,\widetilde{M}$. We refer to these contributions as perturbative, denoting them with a $p$ subscript. The $d_k$ and $d^D_k$ are known instanton coefficients calculated indirectly in \cite{Seiberg:1994bz, Ito:1995ga} and directly in \cite{Nekrasov:2002qd}. 
We computed their explicit values up to 25 and 30 instantons respectively using the method of \cite{Tachikawa:2013kta, Chan:1999gj}. We demonstrate the convergence of these instanton expansions in Appendix~\ref{app:convergence}.

We have seen that the IR dynamics of Seiberg Witten theory is encapsulated in the analytic structure of the function $A(u),\,A_D(u)$ given in \eqref{eq:AADu} (in fact, only from their monodromies around their branch points). Furthermore, this information is preserved under \textit{EM duality transformations} in the duality group $SL(2,Z)$. 
Two transformations, $S$ and $T$, generate $SL(2,Z)$. 
$S$ corresponds to taking $a=d=0$, $b=1$, and $c=-1$, and thus implements the textbook symmetry of electromagnetism that interchanges the electric and magnetic fields: $\vb{E} \to \vb{B}$, $\vb{B}\to -\vb{E}$. $T$ corresponds to $a=b=d=1$ and $c=0$ which simply shifts $\theta$ by $2 \pi$ and is well known to be an exact symmetry. 
A general duality transformation is described by a matrix 
\begin{eqnarray}
    \mathcal{M}=\colmatt{a&b\\c&d}\,, \qquad a,b,c,d\in\mathbb{Z} \, ,
\end{eqnarray}
where $ad-bc=1$, under which $(A_D,A)$ transform as $(A_D,A)\rightarrow \mathcal{M}(A_D,A)$. By \eqref{eq:tau}, $\tau$ transforms under duality as
\beq
    \tau^\prime =\frac{a \tau +b}{c \tau +d}~\,,
\label{eq:tautrans}
\eeq
while the charges of BPS states transform as
\beq
    \colvec{g^\prime\\q^\prime} = \mathcal{M}^{-1\,T}\colvec{g\\q}\,.
\label{sl2zcharge}
\eeq
Note that duality preserves the masses of the BPS states \eqref{eq: BPS mass} regardless of our identification of the axion in the low energy EFT of SW theory.

\section{The Seiberg-Witten \texorpdfstring{${\cal R}$}{R}-Axion}
%
SW theory has an anomalous $U(1)_{\cal R}$ symmetry, which is spontaneously broken by the VEV and plays the role of a Peccei-Quinn \cite{Peccei:1977hh} symmetry.
In the UV, the phase of the adjoint then plays the role of the (${\cal R}$-) axion.
The ${\cal R}$-axion has been studied in different contexts \cite{Farrar:1982te, Nelson:1995hf, Bagger:1994hh, Komargodski:2009rz, Unwin:2024yqq, Dvali:2024dlb, Dobrescu:2000yn, Carpenter:2009sw, Banks:1993en, Miller:2003hm} including collider phenomenology \cite{Goh:2008xz, Bellazzini:2017neg, Arganda:2018cuz}. 
In the low-energy effective theory there is only\footnote{At two isolated singularities on the moduli space there are additional massless hypermultiplets -- these are BPS monopoles/dyons that become massless at the singularities. 
Moving slightly away from these singularities, these BPS states get a mass and we can integrate them out.} an ${\mathcal N}=2$ gauge multiplet, $A$, whose scalar component has charge $2$ under the IR ${\cal R}$-symmetry \cite{Seiberg:1994bz}.
The IR axion $a(x)$ is then a spatially dependent \textit{phase} of that scalar component of $A$, {\it i.e.}
\begin{equation}\label{eq:axexp}
    A(x)= A^v(u)\,e^{i\frac{a(x)}{f(u)}}\,. 
\end{equation}
where $A^v(u)$ is the (generally complex) VEV of $A$ as a function of the Coulomb branch coordinate $u$, and $f(u) = \sqrt{2} |A^v(u)| / e(u)$ is the axion decay constant, chosen so that the axion kinetic term is canonically normalized\footnote{In principle we could have also expanded in the radial fluctuation of $A$ about its VEV, but it is decoupled from the axion and irrelevant for the present discussion.}.
To reduce notational clutter, for the rest of the discussion we will suppress the explicit $u$ dependence of $A^v,\,f,\,e,\,$ and $\theta$.

First note that the SW solution fixes
\begin{eqnarray}\label{eq:tau electric frame}
    &&\tau = {\frac{4 \pi i}{e^2_{p}}}-\frac{8\alpha}{2\pi} - \frac{1}{2\pi}\,G\left(\alpha\right)\\[5pt]  &&G(\alpha) \equiv \sum_{k=1}^\infty \, (\tilde{b}_k - i \tilde{c}_k) \,\left|\frac{\Lambda}{A^v}\right|^{4 k} \, \left[\sin\left(4k \alpha \right)+i\cos\left(4k \alpha\right)\right]\,, \nonumber
\end{eqnarray} 
where $\alpha\equiv\frac{a}{f}-\frac{\theta_p}{8}$,
the coefficients $\tilde{b}_k - i \tilde{c}_k = (4k-1)(4k-2)\,d_k$ are related to the instanton coefficients $d_k$ in \eqref{eq:prepotential}, $e_p^2= \pi^2 / \log(2 \abs{A^v} / 3 \Lambda)$ is the perturbative contribution to the coupling, and $\theta_p = - 8 \arg{A^v}$ is the perturbative theta angle.
The coefficients $b_k$ and $c_k$ from Eq.~\eqref{eq:axion coupling} are related to $\tilde{b}_k$ and $\tilde{c}_k$ by a rescaling of $(\Lambda / A^v)^{4k}$.
From the expression \eqref{eq:tau electric frame} for $\tau$ we learn that it has: (a) a real term which is linear in the axion -- this will give the perturbative part of the axion coupling; as well as (b) real and imaginary terms that are trigonometric in the axion. These constitute an explicit breaking of the axion shift symmetry by the instanton corrections, and are automatically consistent with the $a\rightarrow a+2\pi f$ shift of the axion. The attentive reader may note that the real part of $\tau$ couples to $F\widetilde{F}$ in the Lagrangian, while the imaginary part couples to $F^2$. To see this explicitly, we plug in the expansion \eqref{eq:tau electric frame} in the effective Lagrangian \eqref{eq:SWLagrangian} and keep the terms relevant for the axion-photon coupling
\begin{eqnarray}\label{eq:axLag}
    &&{\mathcal L}_{IR}\supset\nonumber\\[5pt]
    &&
    \left(-\frac{1}{2e^2_{p}}+\frac{{\rm Im} \,G(\alpha)}{16\pi^2}\right)\,F_{\mu\nu}F^{\mu\nu}+\frac{8\alpha + {\rm Re}\, G(\alpha)}{16\pi^2}\,F_{\mu\nu}\widetilde{F}^{\mu\nu}\,,\nonumber\\
\end{eqnarray}
where the instanton corrections are encapsulated in the $G(\alpha)$ function (see Section~\ref{sec:instantons} for more details about the generation of these terms).

In these expressions the vacuum $\theta$ angle is kept for generality, and to manifest EM duality. As usual in theories with axions and massless fermions, $\theta$ is not physical, since ${\rm Re}(\tau)$ can always be set to zero by a $U(1)_{\cal R}$ transformation\footnote{To see this note that ${\rm Re}(\tau)$ depends only on $\alpha$, the instanton coefficients $d_k$ are all real (see Eq.~\eqref{eq:coefs} for the first few), and that the axion potential is zero, meaning that $a \rightarrow (\theta_p f / 8) + a$ sets ${\rm Re}(\tau) = 0$.}.

\begin{figure}[t]
    \begin{center}
    \includegraphics[width=0.9\linewidth]{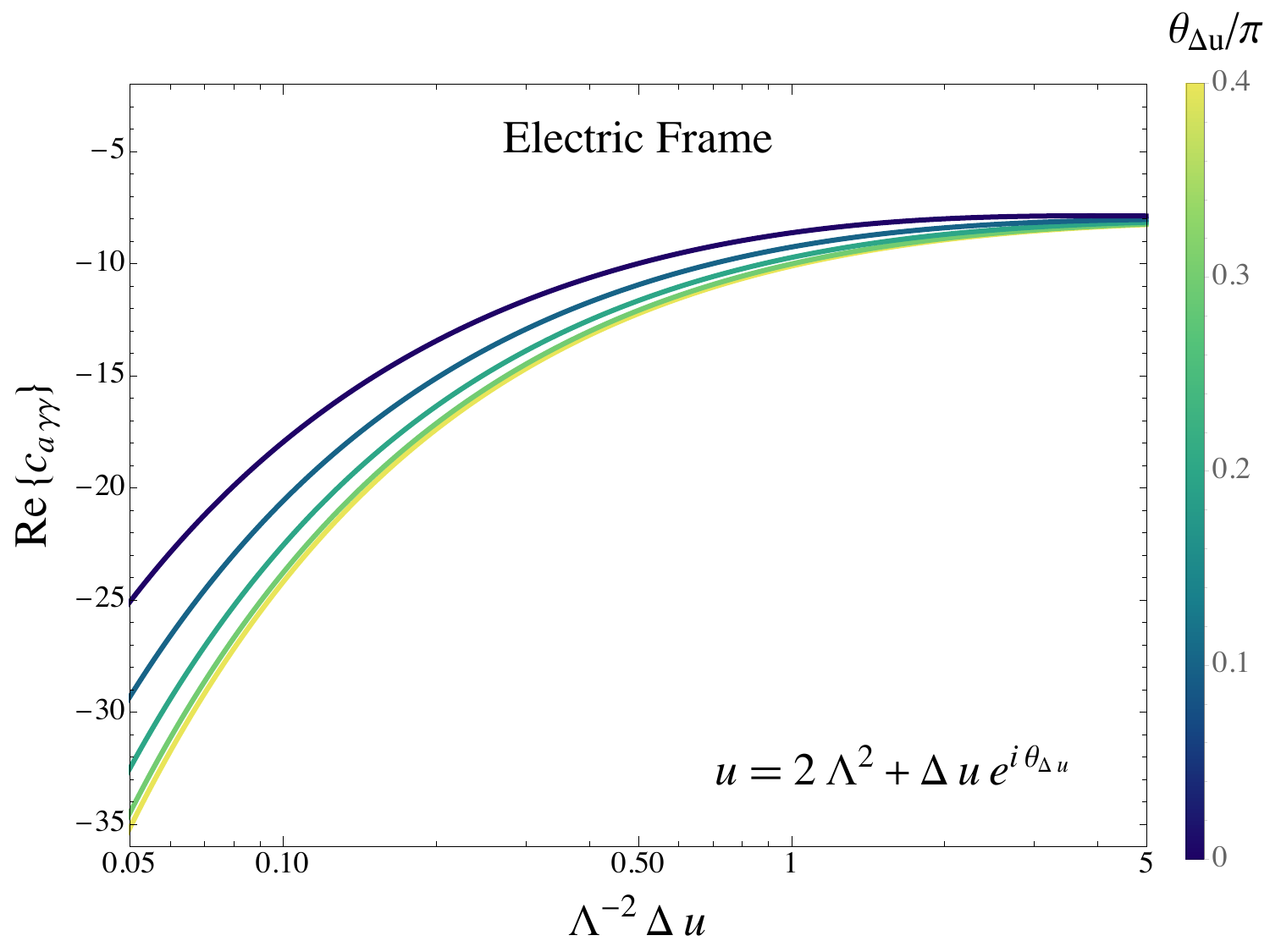}\\
    \includegraphics[width=0.9\linewidth]{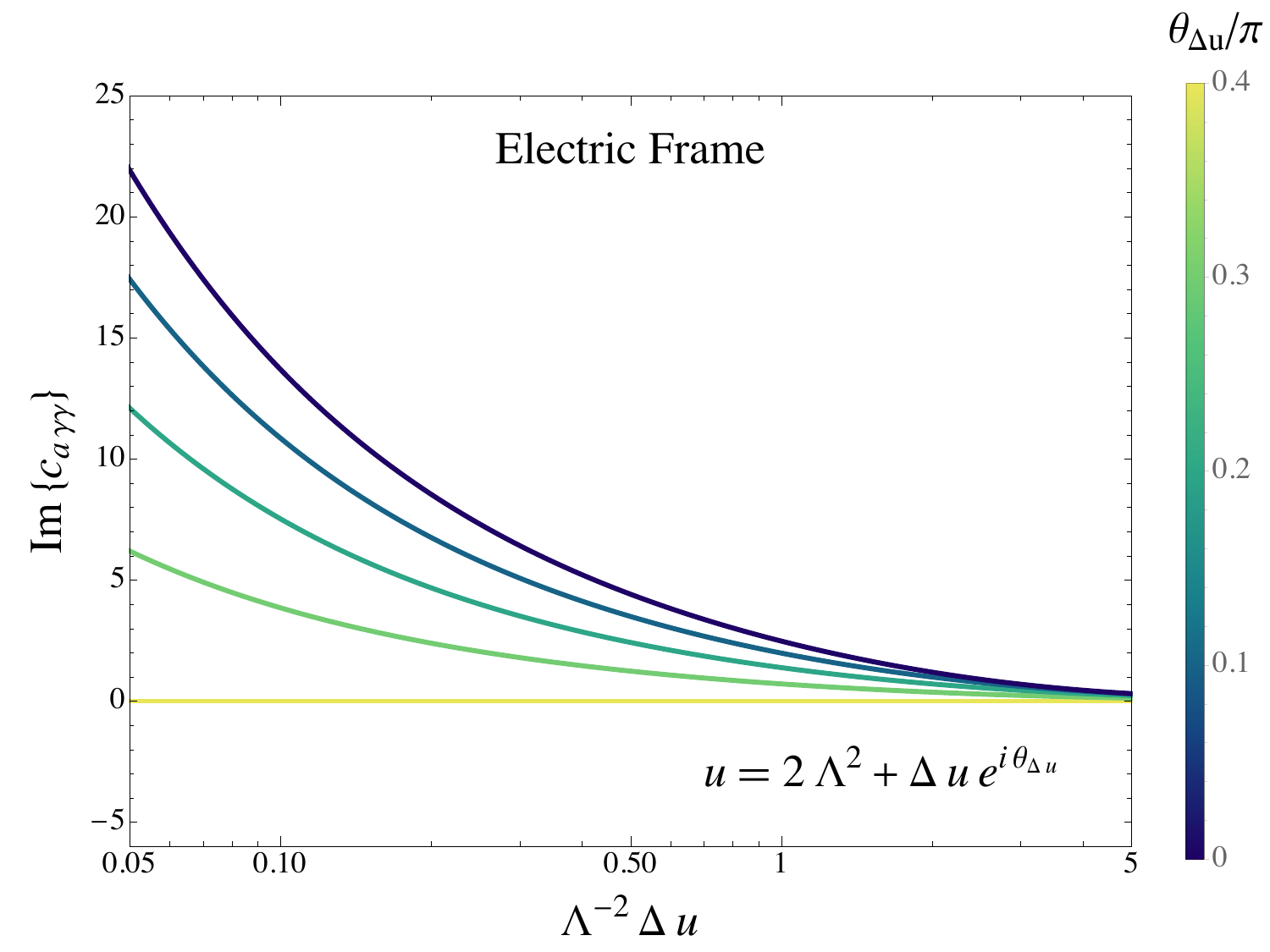}
    \caption{The real and imaginary part of the axion coupling coefficient $c_{a\gamma \gamma}$ in the electric duality frame. The coupling is plotted as a function of the distance $\Delta u$ from the monopole singularity, at different angles in the complex $u$ plane. Evidently, ${\rm Re} \Bqty{ c_{a \gamma \gamma} } \rightarrow -8$ and ${\rm Im} \Bqty{c_{a \gamma \gamma}} \rightarrow 0$ at weak coupling $\Delta u\rightarrow\infty$. ${\rm Re}\Bqty{c_{a \gamma \gamma}}$ diverges at the monopole point $\Delta u=0$. 
    Similarly ${\rm Im}\Bqty{c_{a \gamma \gamma}}$ diverges at the monopole point for $\theta_{\Delta u} \neq 0$.
    }\label{fig:ElectricAxion}
    \end{center}
\end{figure}

Expanding $\tau$ around $a=0$ we find the canonically normalized\footnote{We choose to work in units where electric and magnetic charges are integers. In these units the photon kinetic term is normalized by $1/2$.} IR axion EFT Lagrangian
\begin{eqnarray}\label{eq:electric axion EFT Lag}
    {\cal L}_{IR} & \supset & -\frac{1}{2} \partial_\mu a \, \partial^\mu a - \frac{1}{16\pi} {\rm Im}\Bqty{\bqty{e^2 \tau}_{a = 0} ( F^{\mu\nu} + i \widetilde{F}^{\mu\nu})^2} \nonumber \\[5pt]
    & & - \frac{a}{16 \pi} {\rm Im}
    \Bqty{\bqty{i \frac{e^3}{\sqrt{2}} \pdv{\tau}{A}}_{a = 0} ( F^{\mu\nu} + i \widetilde{F}^{\mu\nu})^2 } + ... \, , \nonumber \\[5pt]
\end{eqnarray}
where the derivative of $\tau$ is with respect to the holomorphically normalized field. We define the complex axion coupling as
\begin{eqnarray}\label{eq:complex axion coupling}
    \frac{g_{a \gamma  \gamma}}{4} = \frac{e^2}{16\pi^2 f} c_{a \gamma \gamma} = \frac{i}{\sqrt{2}} \frac{e^3}{8 \pi} \pdv{\tau}{A} \, ,
\end{eqnarray}
where ${\rm Re}\Bqty{c_{a \gamma \gamma}}$ is identified with the  coefficient of $(e^2 / 16 \pi^2 f)  a F \widetilde{F}$ and ${\rm Re} \Bqty{g_{a\gamma \gamma}}$ the coefficient of $(1/4) a F \widetilde{F}$, both ubiquitous in the literature.
Since we have not set $\theta=0$ we have the additional parity violating $a F^2$ coupling, which generically appears in the one-loop anomaly of dyonic states \cite{Csaki:2010rv}. 
In terms of the prepotential instanton expansion we can explicitly identify the perturbative and non-perturbative contributions to the axion coupling
\begin{eqnarray}\label{eq:coup}
    c_{a\gamma\gamma} = - 8 - \sum_{k=1}^{\infty} 4 k (\tilde{b}_k - i \tilde{c}_k) \,\pqty{ \frac{\Lambda}{A^v} }^{4 k} \, .
\end{eqnarray}
The $-8$ in \eqref{eq:coup} is the perturbative anomaly coming from the four gauginos in $W_\pm$ which have electric charge $2$, whereas the instantons account for the contributions from the BPS states. In Fig.~\ref{fig:ElectricAxion} we present the values of $c_{a \gamma \gamma}$. 
We find that the coupling diverges near the monopole point where $u \rightarrow 2 \Lambda^2$. 
As we will show in more detail in Section~\ref{sec:axion decay rate} this divergence is associated with the anomaly contribution from the strongly coupled magnetic monopole of mass $\sim \sqrt{u - 2\Lambda^2}$.

\section{The \texorpdfstring{$S$}{S}-dual Axion}
\begin{figure}[t]
    \begin{center}
    \includegraphics[width=0.9\linewidth]{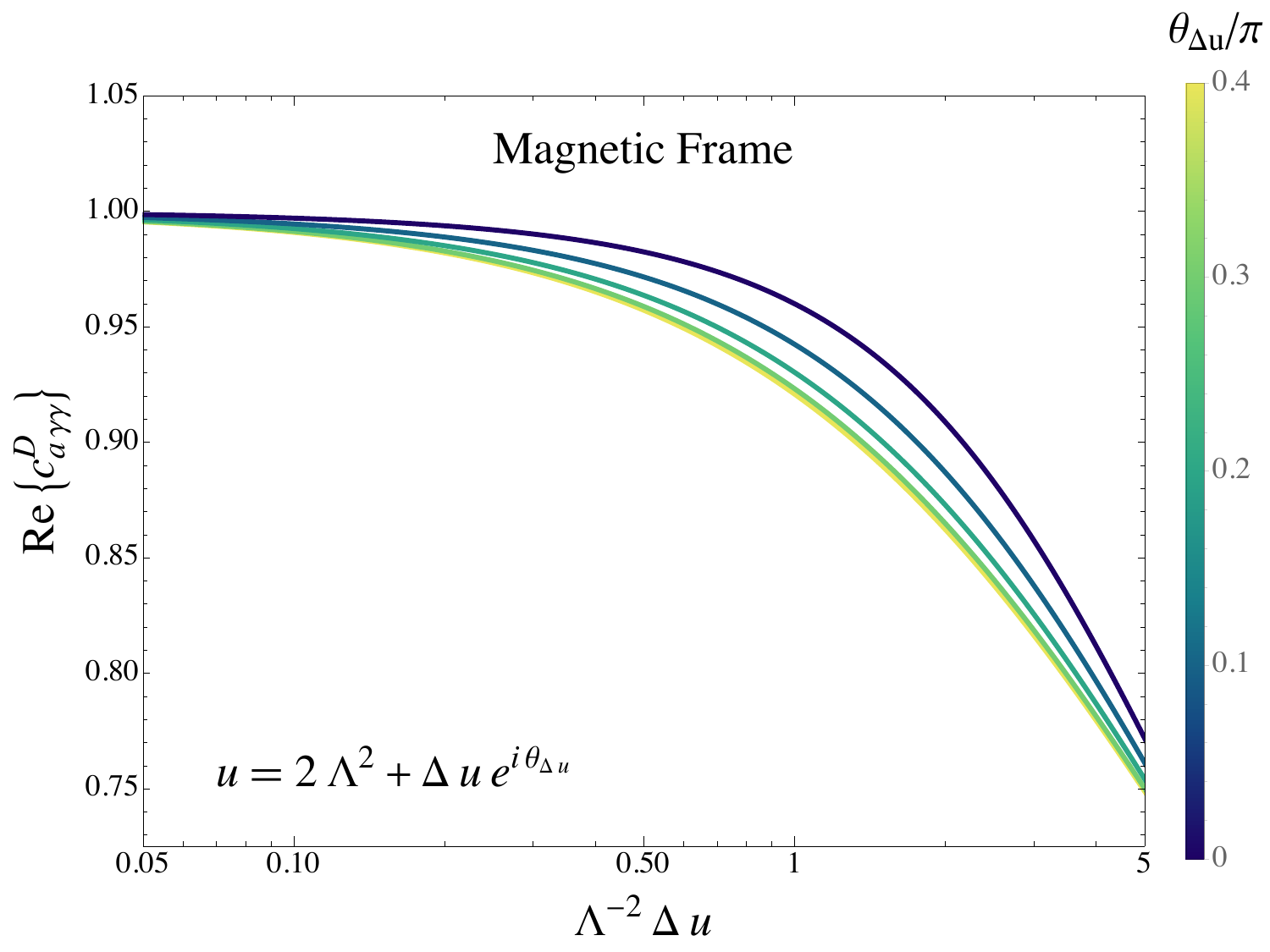}\\
    \includegraphics[width=0.9\linewidth]{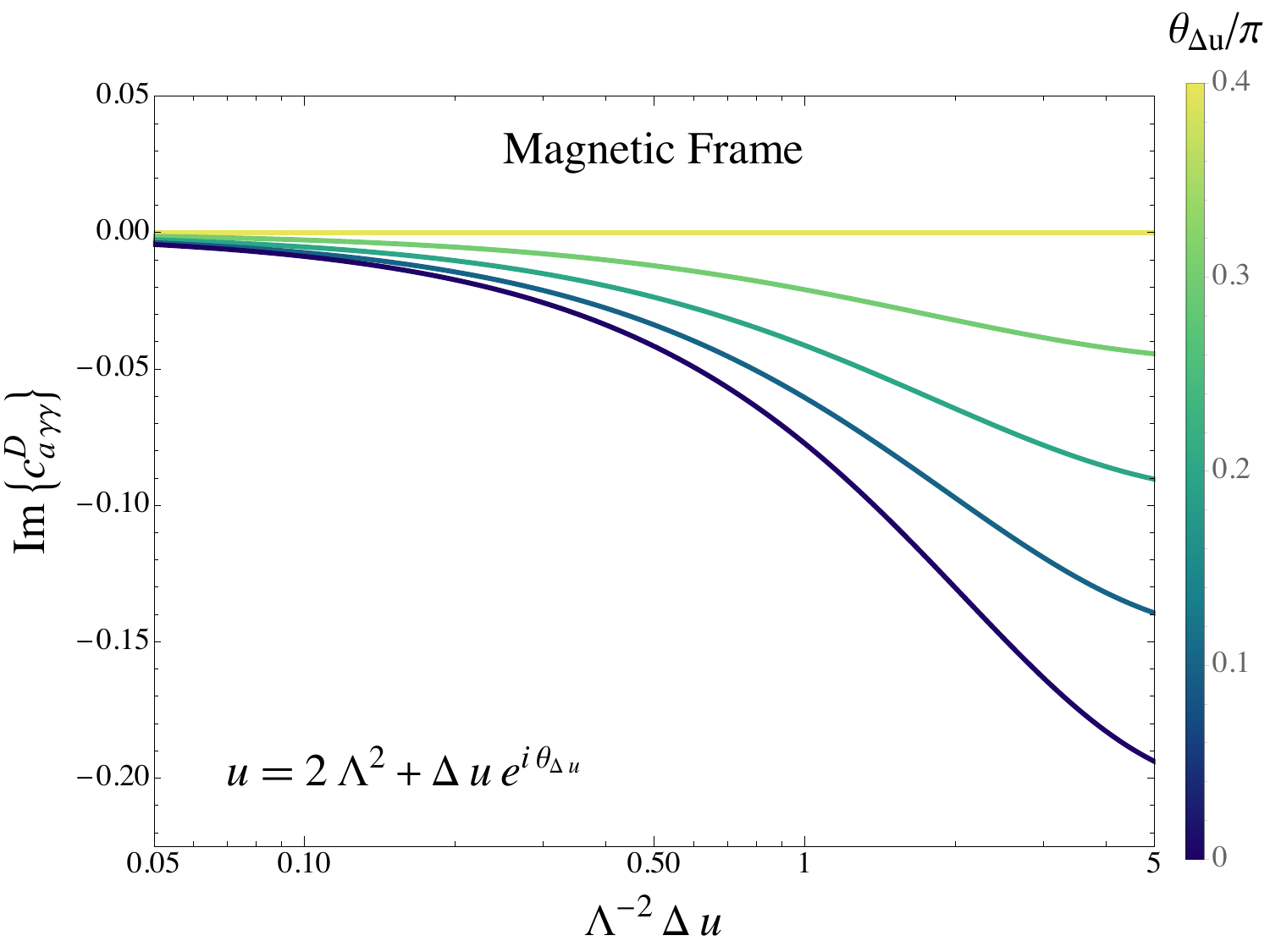}\\
    \caption{The real and imaginary parts of the axion coupling coefficient $c_{a\gamma \gamma}^D$ in the magnetic duality frame. The factors are plotted as a function of the distance $\Delta u$ from the monopole singularity, at different angles in the complex $u$ plane. Evidently, ${\rm Re}\Bqty{c_{a\gamma\gamma}^D} \rightarrow 1$ and ${\rm Im} \Bqty{c_{a\gamma\gamma}^D} \rightarrow 0$ at the monopole point $\Delta u=0$, and significantly deviates from these values at weak coupling. 
    }\label{fig:MagneticAxion}
    \end{center}
\end{figure}

%
To understand this divergence better we can go to the $S$-dual description where the monopole is perturbatively coupled to the photon, referred to as the \textit{magnetic duality frame}. The axion is once again the phase of the scalar superpartner of the photon $A_D$, and its coupling can be discerned from the prepotential ${\cal F}_D$. 
We emphasize that this identification of the axion is uniquely determined by the anomalous $U(1)_{\cal R}$ symmetry.
In other words, the axion transforms under duality.
To better differentiate between fields defined in the two frames, in this paper we use the naming convention of the \textit{electric duality frame}, i.e. we put a ``$D$" subscript on the axion, photon and superpotential of the magnetic frame. Expanding $A_D(x)$ as 
\begin{equation}\label{eq:axexp dual}
    A_D(x) = i A^v_D\,e^{i\frac{a_D(x)}{f_D}}\,,
\end{equation}
where $f_D=\sqrt{2}|A^v_D|/e_D$ and the factor of $i$ is chosen such that $A^v_D(u)$ is positive and real for $u > 2$ along the real line.
Repeating the same procedure as before we find the effective Lagrangian
\begin{eqnarray}\label{eq:magentic axion EFT Lag}
    {\cal L}^D_{IR} & \supset & -\frac{1}{2} \partial_\mu a_D \partial^\mu a_D - \frac{1}{16 \pi} {\rm Im}\Bqty{\bqty{e_D^2 \tau_D}_{a_D = 0} ( F_D^{\mu\nu} + i \widetilde{F}_D^{\mu\nu})^2} \nonumber \\[5pt]
    & & - \frac{a_D}{16 \pi} {\rm Im}
    \Bqty{ \bqty{-\frac{e_D^3}{\sqrt{2}} \pdv{\tau_D}{A_D}}_{a_D = 0} ( F_D^{\mu\nu} + i \widetilde{F}_D^{\mu\nu})^2 } + ... \, , \nonumber \\[5pt]
\end{eqnarray}
and the magnetic frame analog axion couplings
\begin{eqnarray}
    \frac{g_{a \gamma \gamma}^D}{4}  & = &  \frac{e_D^2}{16\pi^2 f_D} c_{a \gamma \gamma}^D = -\frac{1}{\sqrt{2}} \frac{e_D^3}{8\pi}\pdv{\tau_D}{A_D} \, , \nonumber\\[5pt]
    c_{a\gamma \gamma}^D & = & 1 + \sum_{k=1}^\infty \frac{k}{2} \pqty{\tilde{b}_k^D - i \tilde{c}_k^D} \pqty{\frac{A_D^v}{\Lambda}}^k
\end{eqnarray}
where the coefficients $\tilde{b}_k^D - i \tilde{c}_k^D = (k+1)(k+2)d_k^D$ are determined by the instanton coefficients in \eqref{eq:prepotential} and $\theta_{D,p}={\rm arg}(A^v_D)-\frac{\pi}{2}$. Since in this frame the monopole is \emph{weakly} coupled the constant term in $c^D_{a \gamma\gamma}$ is interpreted as the perturbative chiral anomaly from the magnetic monopole, whereas the instanton contributions describe contributions from the gauginos and other BPS states. 
In Fig.~\ref{fig:MagneticAxion} we present the axion coupling $c_{a \gamma \gamma}^D$ in the magnetic frame.
We find that the dominant contribution is the perturbative coupling from the monopole.
As we will subsequently show in Section~\ref{sec:axion decay rate} the seeming inconsistency between $c_{a\gamma\gamma}$ and $c_{a\gamma\gamma}^D$ is resolved by the different prefactors $e^2/f$ and $e_D^2/f_D$ in the two frames. 

\section{Duality Invariance and \texorpdfstring{$a\rightarrow \gamma\gamma$}{a -> gamma gamma} Decay Rate}\label{sec:axion decay rate}
%
The two Lagrangians \eqref{eq:electric axion EFT Lag} and \eqref{eq:magentic axion EFT Lag} provide two equivalent (in this case $S$-dual) descriptions of the same IR physics. 
Therefore, we expect that all canonically normalized physical observables computed from these equivalent Lagrangians agree. In particular, this implies that canonically normalized amplitudes should be identical up to a phase~\cite{Colwell:2015wna}.
In this section we will prove analytically that this is the case for the rate for $a\rightarrow \gamma\gamma$, which is of central importance in axion physics. Moreover we will show that this rate becomes \textit{large} close to the monopole singularity, in stark contrast with the standard PQ scenario. Analyzing the situation in both the electric and magnetic duality frame will provide a complete understanding of why this rate becomes large.

\begin{figure}[t]
    \begin{center}
    \includegraphics[width=0.9\linewidth]{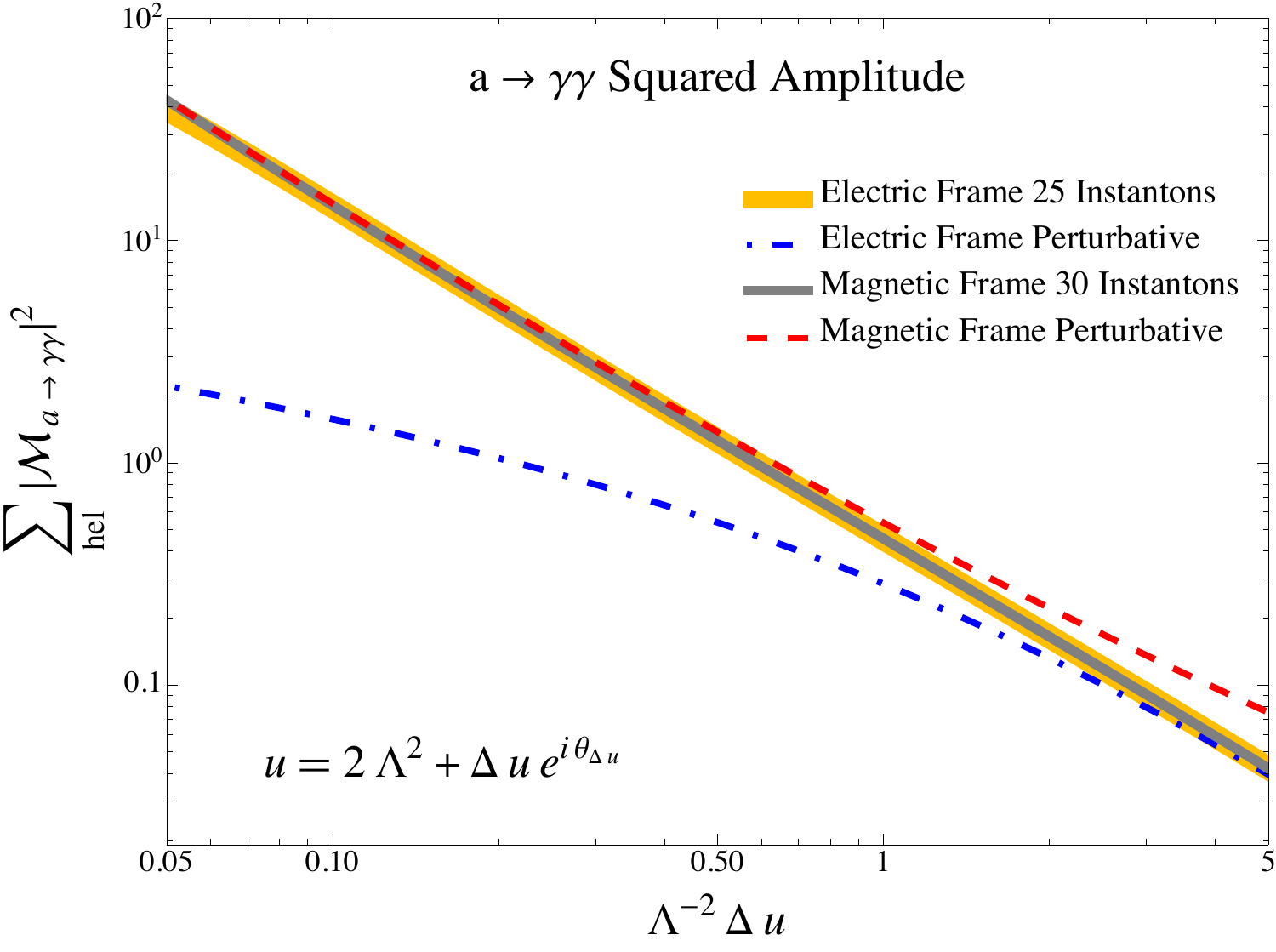}
    \caption{The duality-invariant sum $|\mathcal{M}_{++}|^2+|\mathcal{M}_{+-}|^2$ as calculated in different frames. Orange: calculated from ${\cal \tau}(A)$ with $25$ instanton terms. Gray: calculated from ${\cal \tau}_D(A_D)$ with $30$ instanton terms. Blue: keeping only the perturbative contribution to ${\cal \tau}(A)$. Red: keeping only the perturbative contribution to ${\cal \tau}_D(A_D)$. Near the monopole point we see that the non-perturbative instanton contributions in the electric frame sum up exactly to the perturbative contribution in the magnetic frame.
    }\label{fig:agammagamma}
    \end{center}
\end{figure}

Note that in the presence of a non-zero vacuum theta angle $\theta$, the axion couples also to $FF$ in \eqref{eq:axLag} (and similarly to $F_DF_D$ in the magnetic frame). Thus, when we compute the rate for $a\rightarrow \gamma\gamma$, we have to sum over the squares of the amplitudes $\mathcal{M}_{+-,-+}$ generated by $aF\widetilde{F}$ and $\mathcal{M}_{++,--}$ generated by $aFF$. Altogether the rate is
\begin{eqnarray}\label{eq:rate}
    \Gamma^{\rm el}_{a\rightarrow \gamma\gamma}&\propto& \sum_{h,h'}|\mathcal{M}_{h h'}|^2 \propto 2\left(|\mathcal{M}_{+-}|^2+|\mathcal{M}_{++}|^2\right) 
    \nonumber\\&& \propto \frac{e^4}{f^2}\,\abs{c_{a\gamma\gamma}}^2 \propto \abs{g_{a \gamma\gamma}}^2 
\end{eqnarray} 
Where we omit all constant numerical factors and the phase space integral, which are irrelevant to prove the duality invariance of the rate. In the magnetic frame we get analogously
\begin{eqnarray}\label{eq:rateD}
    \Gamma^{\rm mag}_{a\rightarrow \gamma\gamma}&\propto& \abs{g^D_{a \gamma \gamma}}^2 \,.
\end{eqnarray} 
The point is that $\abs{g_{a \gamma \gamma}}^2$ is in fact \textit{duality invariant}. Using the canonical relation $\tau = - \tau_D^{-1}$, the Seiberg-Witten relation $\partial A / \partial A_D$ and the holomorphic properties of the prepotential we find
\begin{eqnarray}
    e^3 \pdv{\tau}{A} = \pqty{-\frac{\overline{\tau}_D}{\abs{\tau_D}}}^3 e_D^3 \pdv{\tau_D}{A_D} \, ,
\end{eqnarray}
meaning the amplitudes in the two frames are the same up to a phase (for more detail see Appendix~\ref{app:proof duality invariance}). Consequently, we have
\begin{eqnarray}\label{eq:rateeq}
    \Gamma^{\rm el}_{a\rightarrow \gamma\gamma}=\Gamma^{\rm mag}_{a\rightarrow \gamma\gamma}\,,
\end{eqnarray} 
as had to be the case, since the two frames describe the same physics. 
In Figure~\ref{fig:agammagamma} we plot the duality-invariant sum of a amplitudes $|\mathcal{M}_{+-}|^2+|\mathcal{M}_{++}|^2$ as a function of the Coulomb branch coordinate $u$. 
Remarkably, we see that the squared amplitude diverges as we approach the monopole point. 
This is very different from the usual PQ mechanism! 
Furthermore, we see that near the monopole point the decay rate is dominated by the monopole anomaly contribution, and far from the monopole point the decay rate is dominated by the gaugino anomaly.

We emphasize that the key for the matching between frames is the identification of the axion in the different Eqs.~\eqref{eq:axexp} and \eqref{eq:axexp dual}, which was not a choice but rather uniquely determined by $U(1)_{\cal R}$. We note that with this identification there is no ``dual Witten effect'', since the BPS mass spectrum $\eqref{eq: BPS mass}$ remains invariant under all duality transformations. 

Nevertheless one could define ''non-standard axion electrodynamics" in the spirit of \cite{Heidenreich:2023pbi} by \textit{remaining in the electric frame}, yet performing a non-linear field redefinition $a\rightarrow a_D$. It is perfectly consistent to perform such a nonlinear field redefinition in the path integral, though it does obscure the $\mathcal{N}=2$ supersymmetry of the theory. Furthermore, the $U(1)_R$ that $a_D$ is a pseudo-Goldstone of is not linearly realized on the fields of the electric frame. In the field-redifined theory with field content $a_D,\,F^{\mu\nu},\ldots$, one could ask how the mass of the charginos in the electric frame depends on $a_D$. Indeed, the chargino mass is $M=|A|$, which depends nontrivially on $a_D$, in what Ref.~\cite{Heidenreich:2023pbi} called a ``dual Witten effect'' of non-standard axion electrodynamics. Note, however, that from the point of view of the ``normal" electric frame axion $a$, fluctuations in $a_D$ are simply a mix of fluctuations in $a$ and the radial mode $|A|$. From that point of view, what the ``dual Witten effect" in $a_D$ really tells us is that fluctuations in the radial mode $A$ are phenomenologically disfavored, and we will not consider them further.

\section{Axion Coupling from Instantons}\label{sec:instantons}
%
The axion-photon coupling receives perturbative and non-perturbative contributions. While the perturbative contribution is readily understood in terms of the triangle anomaly of the $U(1)_{\cal R}$ current, the non-perturbative, $2\pi f$-shift-symmetric contributions originate from instantons in the UV theory which generate terms in the holomorphic pre-potential $\mathcal{F}(A)$. In the following we will sketch how the one-instanton contribution is obtained. We work in the weak-coupling regime far out in moduli space, where a semi-classical instanton expansion is valid and where the low-energy degrees of freedom can be trivially identified with the UV degrees of freedom along the unbroken direction. For more details see~\cite{Seiberg:1988ur,Finnell:1995dr,Dorey:1996hu,Ito:1996np}.

\begin{figure}[t]
    \begin{center}
    \includegraphics[width=0.8\linewidth]{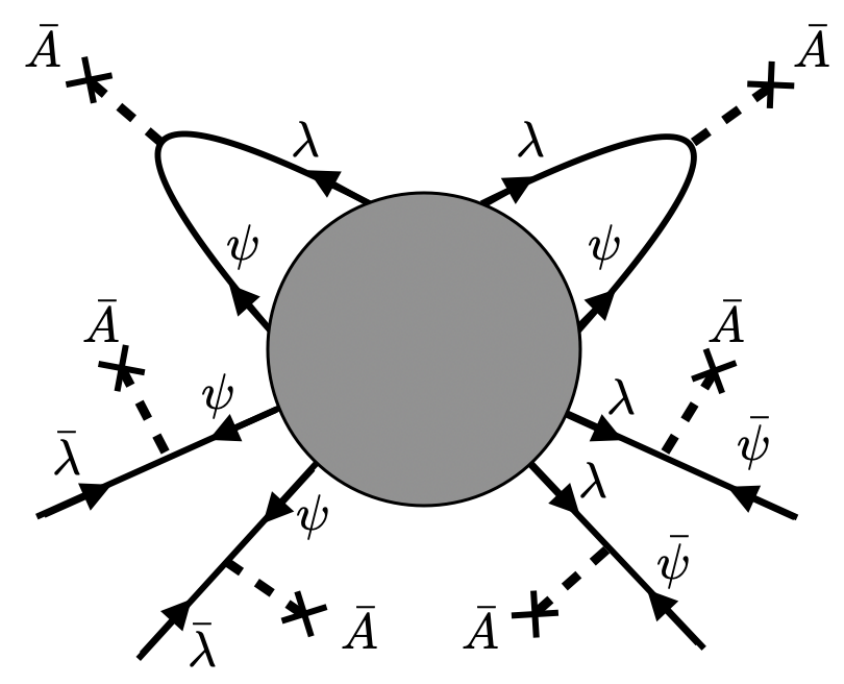}
    \caption{One-instanton contribution to the correlator $\langle \bar{\psi} \bar{\psi} \bar{\lambda}\bar{\lambda} \rangle$ which can be matched to the operator ${\rm Im}\Bqty{ \pqty{ \partial^4 \mathcal{F} / \partial A^4 } \lambda^2 \psi^2}$ which allows us to identify one-instanton generated non-perturbative contributions to the pre-potential. Arrows denote chirality flow and crosses the insertion of the scalar VEV $\bar{A}$.
    }\label{fig:swInstanton}
    \end{center}
\end{figure}

In $\mathcal{N}=1$ language the UV theory contains a vector multiplet $(v^a, \lambda^a, D^a)$ and a chiral multiplet $(A^a, \psi^a, F^a)$, both in the adjoint representation of $SU(2)$. In the instanton background the gauginos and the fermions in the chiral multiplet have four zero modes each, two classical and two supersymmetric/superconformal. When the scalar $A^a$ gets a VEV, which we take to be $A^a \equiv A \delta^{a 3}$, it breaks $SU(2)$ to $U(1)$, what makes the instanton calculation IR finite and lifts the four superconformal zero modes via the Yukawa interaction $i g \epsilon^{abc} \psi^a \lambda^b \bar{A}^c$ and mixes classically massless $\bar{\lambda} \equiv \bar{\lambda}^3$ and $\bar{\psi} \equiv \bar{\psi}^3$ with supersymmetric zero modes $\psi_{\rm SS}$ and $\lambda_{\rm SS}$, respectively. The instanton thus contributes to the $U(1)_{\cal R}$ violating correlator $\langle \bar{\psi} \bar{\psi} \bar{\lambda}\bar{\lambda} \rangle$ via the diagram shown in Figure~\ref{fig:swInstanton}. Using the diagram, the parametric dependence can be easily estimated using power counting rules (see e.g.~\cite{Csaki:2023ziz})
\beq\label{eq:instantonOperator}
\begin{split}
    \langle & \bar{\psi}(x_1) \bar{\psi}(x_2) \bar{\lambda}(x_3)\bar{\lambda} (x_4) \rangle \\ 
    & \propto \int d^4 x_0 \int \frac{d\rho}{\rho^5} (\Lambda \rho)^4 \bar{A}^6 \rho^{12} \,  e^{-4\pi^2 A \bar{A} \rho^2} \prod_{i=1}^4 S_F (x_i - x_0)\\
    & \propto \frac{\Lambda^4}{A^6}\int d^4 x_0 \prod_{i=1}^4 S_F (x_i - x_0)\,,
\end{split}
\eeq
where $S_F (x_i - x_0)$ is a fermion propagator and we left the spinor indices and gauge couplings implicit. Note that the instanton size integral makes the result holomorphic, providing a non-trivial consistency check for its contribution to the pre-potential.

Eq.~\eqref{eq:SWLagrangian} indeed contains a term 
\beq\label{eq:prePotFermion}
\mathcal{L}_{IR} \supset \frac{1}{4\pi} \text{Im}\left[ -\frac{1}{4} \pdv[4]{{\cal F}}{A} \psi^2 \lambda^2 \right]\,,
\eeq
which induces a fermionic correlator equivalent to Eq.~\eqref{eq:instantonOperator}. One can match the two correlators to determine $d_1$, the coefficient of the $A^2 (\Lambda/A)^4$ term in $\mathcal{F}(A)$. Carefully accounting all coefficients one finds~\cite{Finnell:1995dr} $d_k = 1/2$, what agrees with the determination from the analytic solution of Seiberg-Witten theory in Eq.~\eqref{eq:coefs}, what establishes that the first non-perturbative coefficient in $\mathcal{F} (A)$ is generated by a one-instanton configuration in the UV theory. Note that the fermion correlator in Eq.~\eqref{eq:instantonOperator} is related to a bosonic correlator, also generated by instantons, containing the axion and two photons through a supersymmetry transformation. However, practically it is simpler to obtain the fermion correlator, as done above.

\section{Conclusions}
In this paper we gave a concrete proof-of-principle showing that axion couplings to photons need not be quantized, as it is often assumed, or even small. The additional contributions originate from non-perturbative effects due to monopoles (and other BPS states), which can also be interpreted as a sum of instanton corrections. These terms are periodic functions of the axion hence the coupling quantization arguments do not apply to them. Theories with magnetic monopoles, like Seiberg-Witten, give a concrete realization of the issues involved and are consistent with the one-loop-exact holomorphic calculation of the anomaly given in ref. \cite{Csaki:2010rv}. Our results have many implications for axion searches that we hope to follow up in future work. What are the correct equations of motion for axion electrodynamics when the Bianchi identity is not imposed? What is a safe and self-consistent way to account for these new contributions to the axion coupling when we do not have a non-perturbative calculation of the low-energy effective theory?

\begin{acknowledgments}
We are grateful to  Hsin-Chia Cheng, Ben Heidenreich, Liam McAllister, Jacob McNamara, Matt Reece and David Tong for helpful discussions and their illuminating comments on the draft. We are especially grateful to Shimon Yankielowicz for suggesting to demonstrate our claims in an exactly solvable model. 
This work was initiated and performed in part at the Aspen Center for Physics, which is supported by National Science Foundation grant PHY-2210452.
CC is supported in part by the NSF grant PHY-2309456 and in part by the US-Israeli BSF grant 2016153. 
RO acknowledges partial support from the ERC Starting Grant “Light-Dark” (Grant No. 101040019), BSF Travel Grant No. 3083000028, the Milner Fellowship, and the gracious hospitality of Cornell University.
MR was supported by NSF Grant PHY-2310429, Simons Investigator Award No.~824870, DOE HEP QuantISED award \#100495, the Gordon and Betty Moore Foundation Grant GBMF7946, and the U.S.~Department of Energy (DOE), Office of Science, National Quantum Information Science Research Centers, Superconducting Quantum Materials and Systems Center (SQMS) under contract No.~DEAC02-07CH11359. 
The work of OT is supported in part by the NSF-BSF Physics grant No 2022713. 
JT was supported in part by the DOE under grant DE-SC-000999. 
This project has received funding from the European Research Council (ERC) under the European Union’s Horizon Europe research and innovation programme (grant agreement No. 101040019).  Views and opinions expressed are however those of the author(s) only and do not necessarily reflect those of the European Union. The European Union cannot be held responsible for them.

\end{acknowledgments}

\appendix

\begin{figure}[ht!]
    \begin{center}
    \includegraphics[width=0.9\linewidth]{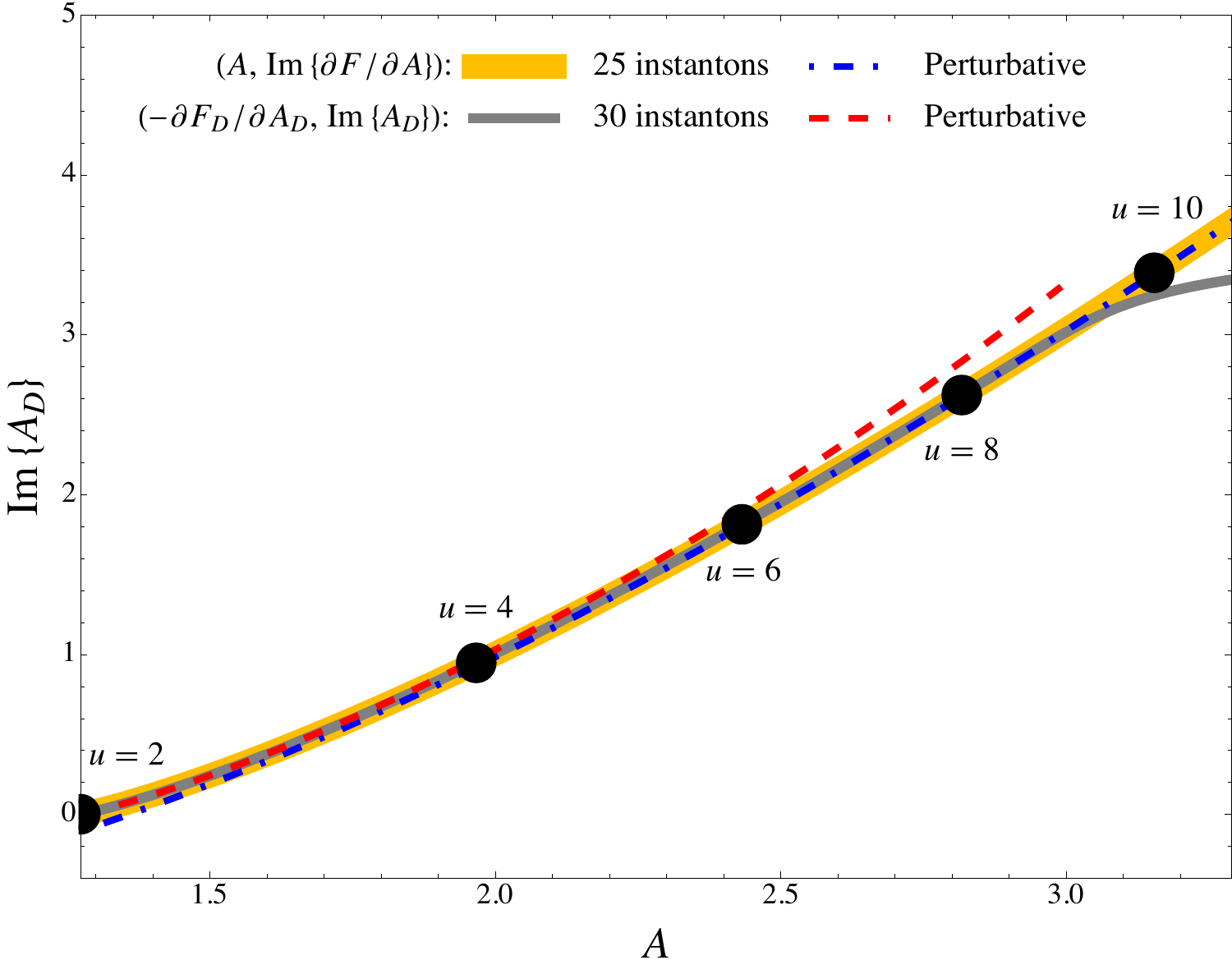}
    \caption{$A_D$ vs. $A$. Orange: calculated from ${\cal F}(A)$ with $25$ instanton coefficients. Gray: calculated from ${\cal F}_D(A_D)$ with $30$ instanton coefficients. Blue: keeping only the perturbative contribution to ${\cal F}(A)$. Red: keeping only the perturbative contribution to ${\cal F}_D(A_D)$. The black circles are $(A(u), A_D(u))$ sample values of $u$, as calculated from \eqref{eq:AADu}.
    We see agreement between the instanton expansions of ${\cal F}$ and ${\cal F}_D$.
    The deviation around $u \sim 10$ is attributed to the loss of validity of this finite term asymptotic series and would be alleviated by considering additional instanton terms. 
    Remarkably, we find that the perturbative one-loop pieces of both expansions are the main contributions in the segment $2 \leq u \leq 10$ along the real axis of the moduli space.
    }\label{fig:conv}
    \end{center}
\end{figure}

\section{Instanton Expansion Convergence}\label{app:convergence}
%
To demonstrate the convergence of the instanton expansions for the prepotential ${\cal F}(A)$ and its Legendre transform  ${\cal F}_D(A_D)$ given in Eq.~\eqref{eq:prepotential}. 
We explicitly show that both expansions yield the same curves $(A,A_D)$ when varying $u$. 
In this manuscript we used the first $30$ instantons coefficients recursively calculated using the method of \cite{Chan:1999gj, Tachikawa:2013kta}. 
Here we present the first six coefficients
\begin{eqnarray}\label{eq:coefs}
    &&d_1=\frac{1}{2},\,d_2=\frac{5}{64},\,d_3=\frac{3}{64},\,d_4=\frac{1469}{32768},\,d_5=\frac{4471}{81920},\nonumber\\[5pt]
    &&d_6=\frac{40397}{524288},\,\ldots
\end{eqnarray}
as well as 
\begin{eqnarray}
    &&d^D_1=\frac{i}{16},\,d^D_2=-\frac{5}{1024},\,d^D_3=-\frac{11i}{16384},\,d^D_4=\frac{63}{524288},\nonumber\\[5pt]
    &&\,d^D_5=\frac{527i}{20971520},d^D_6=-\frac{3129}{536870912},\,\ldots
\end{eqnarray}
In Fig.~\ref{fig:conv} we plot $A_D$ vs. $A$ for Seiberg witten theory, calculated from \eqref{eq:AAD} using the explicit expressions \eqref{eq:prepotential}. 
As one can readily see, the instanton expansion converges rapidly for $\partial {\cal F} / \partial A$ in the entire region, and for $\partial {\cal F}_D / \partial A_D$ at $u\lesssim 9\Lambda^2$. 
Furthermore, we see that the perturbative one-loop pieces of ${\cal F}$ and ${\cal F}_D$ are able to approximate this curve quite well along the real axis of the moduli space.

\vspace{5pt}
\section{Duality Invariance of \texorpdfstring{$a \to \gamma \gamma$}{a -> gamma gamma}}\label{app:proof duality invariance}
%
We show explicitly that $e^3 \partial \tau / \partial A$ is a duality invariant quantity. For a holomorphic function $f(z) = u(z) + i v(z)$ where $u(z),\, v(z)$ are real functions the Cauchy Reimann equations imply
\begin{eqnarray}
    && \frac{1}{v^{3/2}} \pdv{u}{z} = - 2 i \pdv{z} \frac{1}{\sqrt{v}} \, , \label{eq:hol id 1} \\[5pt]
    && \pdv{z} \frac{|f|}{\sqrt{v}} = \frac{\overline{f}^2}{\abs{f}} \pdv{z} \frac{1}{\sqrt{v}} \, . \label{eq:hol id 2} 
\end{eqnarray}
From \eqref{eq:hol id 1} and the canonical relation $\tau = -\tau_D^{-1}$ we find
\begin{eqnarray}
    g_{a\gamma \gamma} \propto \frac{e^3}{4 \pi} \pdv{\tau}{A} = \pdv{e}{A} = \pdv{A_D}{A} \pdv{A_D} \frac{\sqrt{4\pi} \abs{\tau_D}}{{\rm Im }\tau_D}\, .
\end{eqnarray}
However note that in Seiberg-Witten theory $\partial A_D / \partial A = - \tau_D^{-1}$ and with the help of \eqref{eq:hol id 2} we find
\begin{eqnarray}
    \pdv{e}{A} = \pqty{-\frac{\tau_D}{\abs{\tau_D}}}^3 \pdv{e_D}{A} \, ,
\end{eqnarray}
and thus
\begin{eqnarray}
    g_{a\gamma\gamma} = \pqty{-\frac{\tau_D}{\abs{\tau_D}}}^3 g^D_{a\gamma\gamma} \, .
\end{eqnarray}
The upshot is that under duality the $a \to \gamma \gamma$ amplitude changes only by a non-physical global phase.

\bibliographystyle{apsrev4-2}
\bibliography{bibliography}

\end{document}